\documentclass{aastex}
\usepackage{emulateapj5,natbib,psfig}

\newcommand{\gtsim}{\mbox
{{\raisebox{-0.4ex}{$\stackrel{>}{{\scriptstyle\sim}}$}}}}

\newcommand{\mc}{\multicolumn}

\slugcomment{accepted by ApJ}

\shorttitle{Doubly lensed hotspots and the logarithmic slope of the potential}

\shortauthors{Blundell et al}

\begin{document}

\title{A doubled double hotspot in J\,0816+5003 and the logarithmic
slope of the lensing potential}

\author{Katherine M.\ Blundell\altaffilmark{1}, Paul L.\
Schechter\altaffilmark{2},  N.\ D.\ Morgan\altaffilmark{2}, 
Matt J.\ Jarvis\altaffilmark{3}, Steve Rawlings\altaffilmark{1},  
\& John L.\ Tonry\altaffilmark{4}}

\altaffiltext{1}{University of Oxford, Astrophysics, Keble Road, Oxford,
OX1 3RH, UK}
\altaffiltext{2}{Department of Physics, Massachusetts Institute of
Technology, 77 Massachusetts Avenue, Cambridge, MA 02139, USA}
\altaffiltext{3}{Centre for Astrophysics, University of Hertfordshire, 
Hatfield, Herts,  AL10 9AB, UK}
\altaffiltext{4}{Institute for Astronomy, University of Hawaii, 2680
Woodlawn Drive, Honolulu, HI 96822, USA}

\begin{abstract}
  We present an analysis of observations of the doubly-lensed double
  hotspot in the giant radio galaxy  J\,0816+5003 from
  MERLIN, MDM, WIYN, WHT, UKIRT and the VLA.  The images of the two
  hotspot components span a factor of two in radius on one side of the
  lensing galaxy at impact parameters of less than 500\,pc.  Hence we
  measure the slope of the lensing potential over a large range in
  radius, made possible by significant improvement in the accuracy of
  registration of the radio and optical frame and higher resolution
  imaging data than previously available.  We also infer the lens and
  source redshifts to be 0.332 and $\gtsim$ 1 respectively.  Purely on
  the basis of lens modelling, and independently of stellar velocity
  dispersion measurements, we find the potential to be very close to
  isothermal.
\end{abstract}


\section{Introduction}
\label{sec:intro}
In contrast to multiply-imaged, compact flat-spectrum sources, 
strongly-lensed extended, steep spectrum sources occupy a distinctly inferior
position in the gravitational lens pantheon.  Yet, in a paper that marks
the beginning of the modern era in gravitational lensing,
\citet{Pre73} argued that radio lobes were the most promising targets
in searching for lensing.  Indeed, examples in their figure\,6
resemble the lens which is the study of this paper, but since that
paper was written, lensed lobes have proven more difficult to
interpret.

Chief among the difficulties is the continuing poor registration
between the radio and optical reference frames \citep{Deu99}.  Radio
lobes tend not to have optical emission that allows their position to
be determined with respect to any lensing galaxy that might be
identified.  As lobes are extended and polarized, they offer more
potential constraints than multiply imaged point sources, but
modelling such emission requires substantial machinery
\citep{Koc89,Ell96,Dye2005,Vegetti2009}. 
For the most part, lensed lobes have been passed
over in favor of their compact flat-spectrum cousins.

We describe in this paper Merlin and VLA observations of the
doubly-lensed double radio hotspot J\,0816+5003 (a.k.a.\ 6C\,0812+501)
discovered both by \citet{Leh01}, using a clever automated strategy
for finding extended radio emission which is gravitationally lensed,
and independently by the Oxford group searching for high redshift
radio galaxies.  The techniques for high redshift radio galaxy
searches include filtering on the basis of steep radio spectra and
small angular size \citep[e.g.\ ][]{Blundell1998,Jar01a,Jar01} in the
course filtering radio source surveys to optimize seaches for high
redshift radio galaxies.  J\,0816+5003 is an example of a candidate
included in the initial search but subsequently excluded by the
filtering criteria.

The lens we study in this paper is the hotspot complex within the
northern lobe of the giant radio galaxy J0816+5003, which has an
angular extent of over an arcminute and is pictured in figure\,1 of
\citet{Leh01}.  Both the associated radio core, a half arcminute away
to the south west, and its optical counterpart are identified,
allowing accurate registration of the lensed hotspot and the lensing
galaxy.  In addition we find, with higher resolution radio imaging
than was available to \citet{Leh01}, that the hotspot has two
components both of which are doubly imaged.  Finally, the lensed
images are asymmetrically located with respect to the lensing galaxy,
permitting determination of the radial profile of the lensing
potential.  In particular, the images of the two sources span a factor
of two in radius on one side of the galaxy at impact parameters of
less than 500\,pc.

This, together with accurate radio and optical astrometric
registration as well as higher resolution radio images than were
available to \citet{Leh01}, enables us to make a pure-lensing
measurement of the slope of the potential of the {\em total mass} of
the lensing galaxy on the scales probed, including the dark matter.
By combining lensing and dynamical analyses, the SLACS team have
investigated this and other properties \citep[e.g.\
][]{Bolton2007,Gavazzi2007,Bolton2008,Barnabe2009}.  The slope of the
potential figures prominently both in measurements of the Hubble
constant and in discussions of the mass profiles of elliptical
galaxies \citep{Che95,Ber99,Rom99,Wil00,Wit00,Coh01}.  There are very
few examples where one can measure the slopes in this way --- some
exceptions include MG\,1654+134 \citep{Koc95}, 0957+561
\citep{Gro96a,Gro96b,Fis97}, MG\,2016+112 \citep{Tre02}, B\,1152+199
\citep{Rus02}, but cf.\ B1933+503 \citep{Coh01} and especially
PG\,1115+080 \citep{Che95,Rom99}.  Some newly found lenses by the
SLACS team have also yielded constraints on the slope, albeit in most
cases by a comparision of stellar velocity dispersions and Einstein
ring radii \citep{Koopmans2006}.

With regard to measurement of $H_0$, a very small fraction of systems
are suitable for measurement of time delays but as mentioned above,
only a small fraction of systems are suitable for measurement of the
slope of the potential.  Since the overlap of these two sets is slim,
it remains important to glean as much as possible about the slopes of
galaxies potentials and extrapolate from the former to the latter
\citep{Koc04}.   The currently accepted value of $H_0$ is reconcilable
with inferred values from the 10 firm time-delay measurements in
gravitational lenses only if dark matter halos are eliminated and a
constant mass-to-light ratio used.

\section{Observations}
\subsection{VLA imaging}
\label{sec:vla}
J\,0816+5003 was observed for 2 minutes on 1996 Dec 06 in the
A-configuration of the Very Large Array (hereafter VLA) at 5\,GHz
under program code AR365 originally as part of an Oxford survey for high
redshift radio galaxies.   The data were reduced using standard
procedures in the AIPS software package.  The primary flux calibrator
was 3C\,286 and the phase calibrator was 0804+499.  The image showing
its lensed northern hotspot structure is shown in
Figure\,\ref{fig:vla_wht}.

\subsection{Merlin imaging}
\label{sec:merlin}
To obtain higher resolution radio imaging than in
Figure\,\ref{fig:vla_wht}, J\,0816+5003 was observed with six antennas
of the Multi-Element Linked Radio INterferometer (hereafter MERLIN) at
1.7\,GHz on 1997 Jun 27 for 10.58 hrs, during which observations of a
nearby phase calibrator, B\,0803+514, were interleaved.  Again, the
data were reduced using standard procedures within AIPS including
standard polarimetric calibration made possible by the extensive
parallactic angle coverage of the phase calibrator throughout 
the 10+ hour observation.

Figures\,\ref{fig:blobby} and \ref{fig:squash} show contours of the
total intensity MERLIN data in different ways as described in the figure captions,
with the highest resolution image having a point spread function with
FWHM of $0.08^{\prime\prime} \times 0.08^{\prime\prime}$.

\psfig{figure=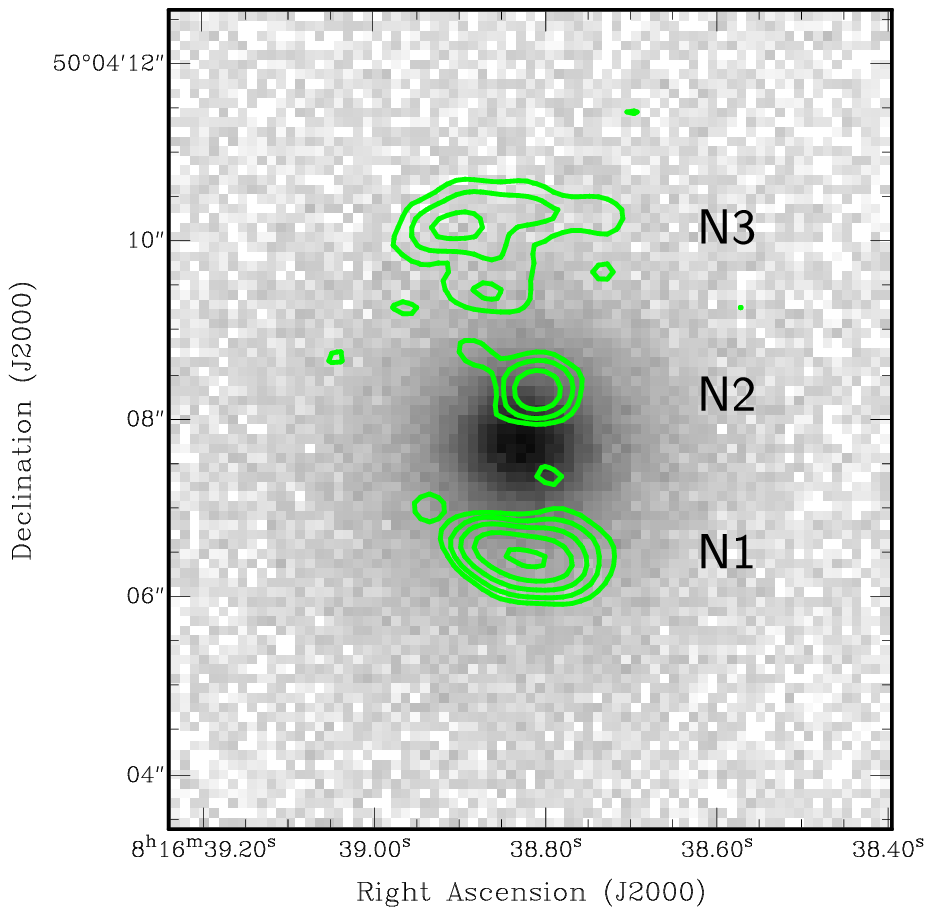,width=9cm,angle=0}
\figcaption{{\label{fig:vla_wht} VLA contours of the north hotspot
    structure of J\,0816+5003 at 5\,GHz. It has a beam of
    0.55$^{\prime\prime}$ $\times$ 0.36$^{\prime\prime}$ at a p.a. of
    82.39\,deg.  The lowest contour is at 0.4\,mJy/beam and adjacent
    contours differ by a factor of $\sqrt{2}$.  The greyscale is an
    $R$-band image from the WHT.  The convention for naming follows
    that instigated by \citet{Leh01}.  }}

\subsection{Optical and near-IR imaging}
\label{sec:optical}

In order to establish the nature of any galaxy associated with
6C\,0816+5003, this source was observed for 5 minutes in $R$-band on
1997 Jan 9 on the William Herschel Telescope (hereafter WHT) using the
Aux Port Camera, and was flat-fielded and reduced using standard
procedures.  A greyscale of this image is presented in
Figure\,\ref{fig:vla_wht}.

Near-infra red K-band observations of J\,0816+5003 were made using the
UFTI instrument for 27 minutes on the United Kingdom Infra-Red
Telescope (hereafter UKIRT) on 2002 Jan 13.  Unfortunately conditions
were not photometric, but approximate flux density boot strapping was
made via a short (9-minute) service observation on 2002 Mar 11.  For
each of these observations, following standard practice, each
successive 60-sec exposure was jittered from the previous position and
a flat-field was created by combining all frames from a particular
observation using a median filter. All data reduction was performed in
IRAF and followed \citet{Wil03}.

\psfig{figure=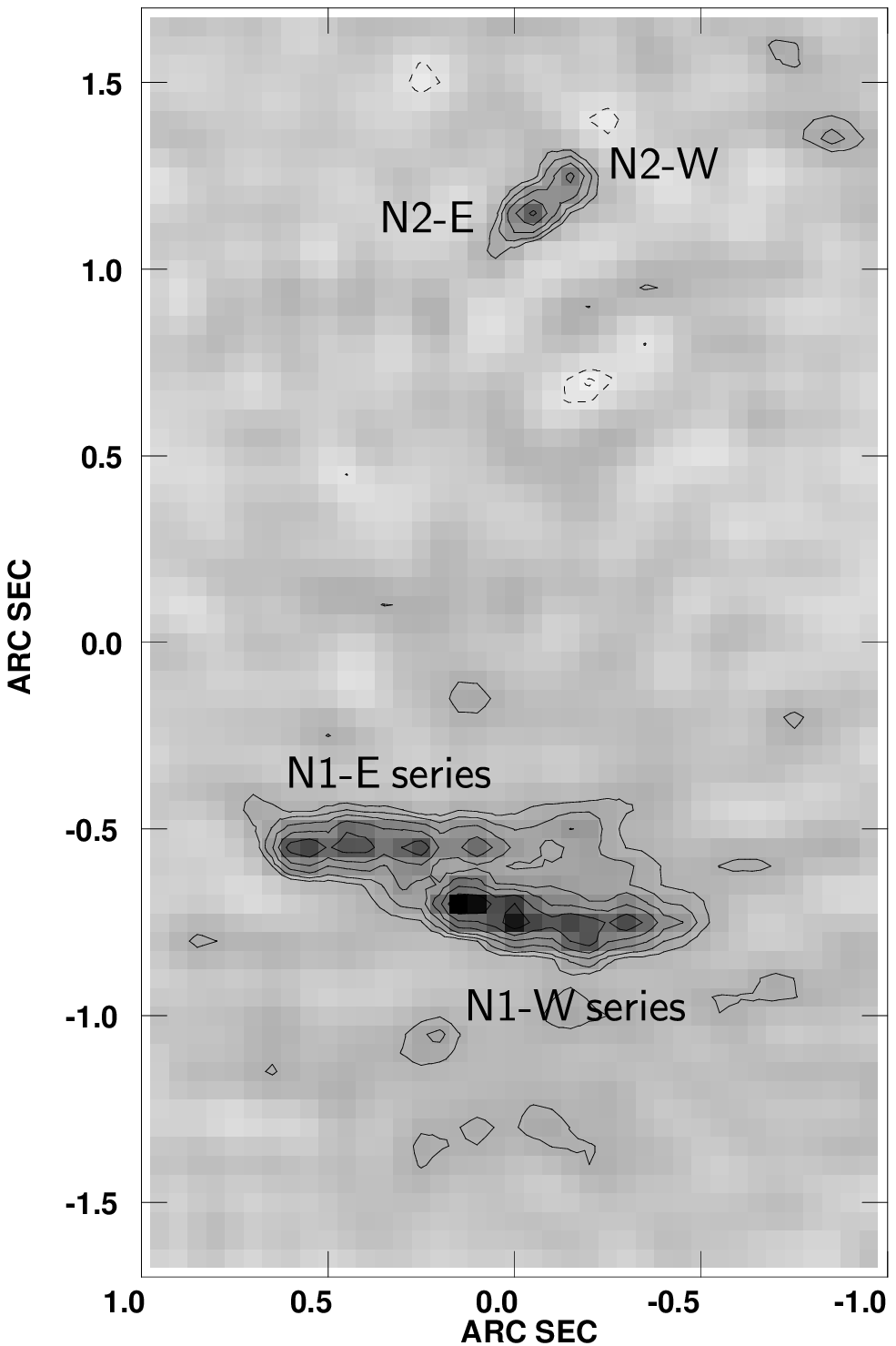,width=8cm} 
\figcaption{{\label{fig:blobby} MERLIN image at 1.7\,GHz of part of
the lensed north hotspot in J\,0816+5003 showing the southern 
two components of the three depicted in Fig\,\ref{fig:vla_wht}.  
This image was made using
the IMAGR routine of the AIPS software package with a Briggs'
robustness parameter of $-3$ and super-resolved with a restoring beam
of $0.08^{\prime\prime} \times 0.08^{\prime\prime}$.   The lowest contour is
0.3\,mJy/beam and adjacent contours differ by a factor of
$\sqrt{2}$. The beam is $0.08^{\prime\prime} \times
0.08^{\prime\prime}$.}}

\psfig{figure=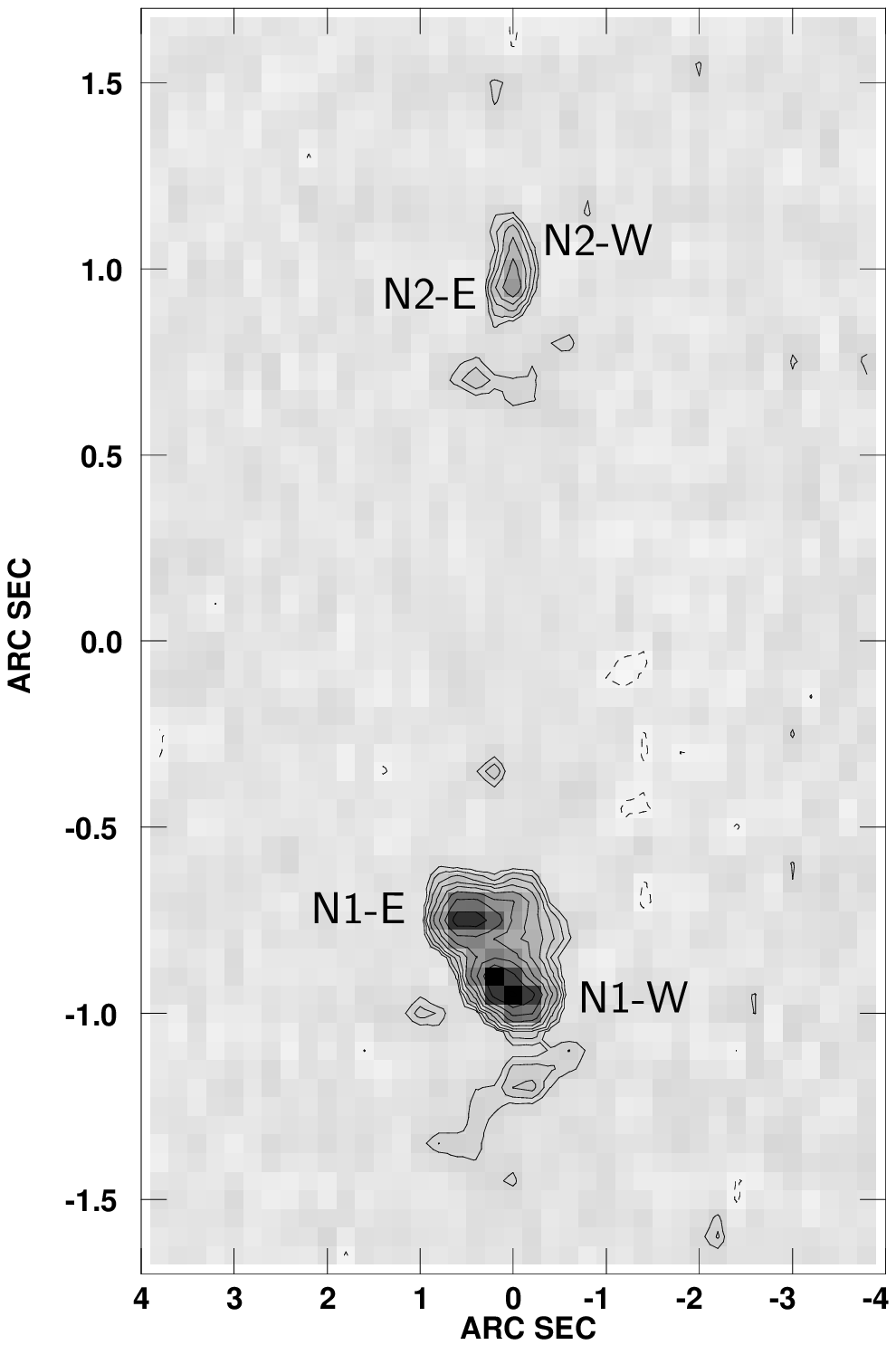,width=8cm} 
\figcaption{{\label{fig:squash}
    MERLIN image at 1.7\,GHz of the same part of the north hotspot
    shown in Fig\,\ref{fig:blobby}, but with the image compressed in
    the East-West direction by a factor of four.  The lowest contour
    is 0.3\,mJy/beam and adjacent contours differ by a factor of
    $\sqrt{2}$. The beam is $0.08^{\prime\prime} \times
    0.08^{\prime\prime}$.  }}

\subsection{Redshifts of lens and lensed galaxy}
\label{sec:specy}

The ISIS spectrograph on the WHT was used on 2002 Feb 5 using the red
and blue arms of the ISIS spectrograph (with R158R and B300B gratings
respectively), in order to find the redshifts of the background
source, namely the radio galaxy, and the lensing galaxy. The spectra
were reduced using standard IRAF procedures. The redshift of the
lensing galaxy is inferred to be 0.332, on the basis of an observed
4000\AA\ break and the Ca H and K absorption lines (at 3968 and 3933
\AA\ respectively) shortward of this.  At this redshift,
$1^{\prime\prime}$ is 4.738\,kpc assuming $H_{\rm 0} = 71 {\rm
  km~s}^{-1}$, $\Omega_{0}=1$ and $\Lambda_{0} = 0.73$.  The
background source, being a radio galaxy, belongs to a relatively
homogeneous category of sources with respect to its host galaxy.  This
is displayed most apparently by the tightness of the radio galaxy K-z
relation \citep{Lil84}. Indeed, with work on the radio galaxy $K$--$z$
Hubble relation \citep{Jar01,Wil03} we are able to make a crude
estimate of the redshift of the radio source.  A magnitude of $K \sim
17.8$ from the UKIRT observations implies a redshift of $z \gtsim
1.0$, which is also consistent with its broad-band optical colours.
We can also gain more information from the optical WHT spectrum, where
the presence of a relatively bright continuum but the absence of any
of the bright forbidden narrow-emission lines, implies that the radio
galaxy probably lies within the redshift desert at $1.2 < z < 1.7$
(i.e.\ [OII] 3727 has been redshifted out of, and Ly$-\alpha$ has yet
to enter, the optical passband).  While the $K$-band magnitude
indicates that the radio galaxy is probably at $z \gtsim 1$, the WHT
spectrum tentatively indicates that $z \sim 1$.  Thus, we can be
certain that the radio galaxy is at a much greater distance than the
lens.

\psfig{figure=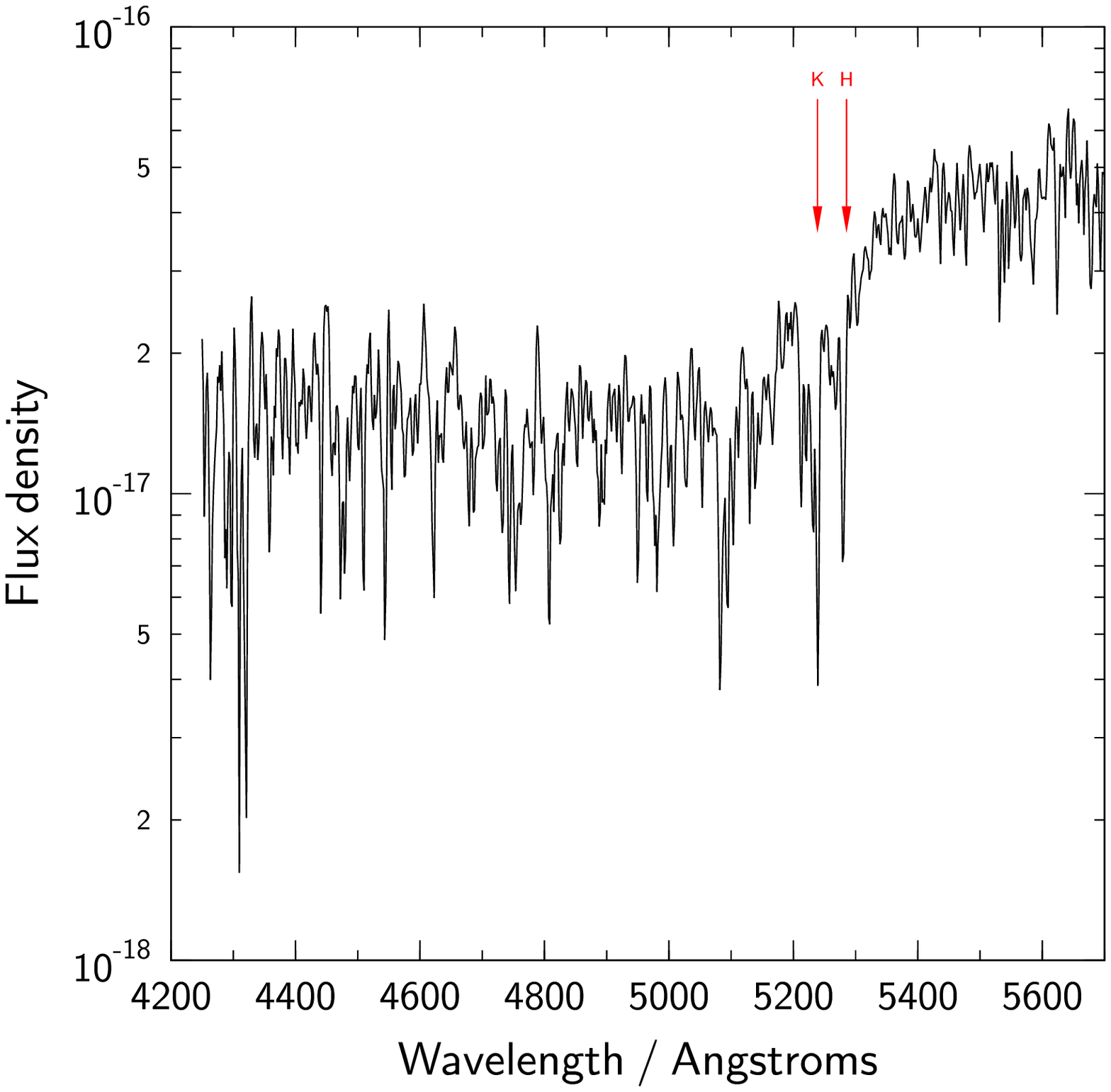,width=8cm,angle=0,bbllx=21 pt,bblly=0
  pt,bburx=500 pt,bbury=520 pt,clip=t} \figcaption{{\label{fig:wht}
    Spectrum from the ISIS spectrograph on the WHT, observed on 2002
    Feb 05.  The redshift of the lens galaxy is inferred to be 0.332,
    on the basis of the H and K absorption lines shortward of the
    4000\,\AA\ break. }}

\section{Optical astrometry}
\label{sec:opticalastrom}

To deduce accurately the slope of the potential of the lensing galaxy
requires secure relative astrometry between the images and the lenses.
In the case of PG\,1115+080, a difference in the image positions of
several parts in a thousand makes a difference between a best-fit
power-law slope of $\alpha = -1$ rather than $\alpha = 0$
\citep{Sch00}, which relates to the surface mass density profile with
radius $r$ as $\Sigma(r) \propto r^{\alpha-2}$.  In the case of
J\,0816+5003, the lensed images are only detectable in the radio
waveband and the lensing galaxies is not visible in the radio.  Thus,
making deductions about the slope of the power-law of the galaxy
lensing its hotspot depends on how well the radio and the
optical/near-IR frames can be tied together.  The derivation of
astrometric solutions using an image from a telescope with reimaging
optics or with a prime focus corrector (such as the WHT, UKIRT or
Keck) and with a reference from the USNO catalog would not be
sufficiently successful, due to off-axis distortions.

A $17^{\prime} \times 17^{\prime}$ $R$-band image of this field was
observed with the Michigan-Dartmouth-MIT 1.3\,m telescope (hereafter
MDM-1.3\,m) by Jules Halpern on 2002 Feb 22.  The astrometry of this
image was fitted with John Thorstensen's automated astrometric
program\footnote{http://www.astro.lsa.umich.edu/obs/mdm/manuals/Manuals/automatch.html}
which matched 156 bright stars over this field from the USNO
catalog\footnote{http://ftp.nofs.navy.mil/projects/pmm/ } and put the
astrometric solution into the fits headers.  This astrometric solution
was transferred to a $4096 \times 2048$ pixel $R$-band image obtained
from the WIYN with 0.5$^{\prime\prime}$ seeing on 2002 Mar 14.

The WIYN and MDM scales were both determined from fits which assumed
that the scale was the same in both directions.  The scale factor in
both cases was determined to better than 1 part in $10^4$.  The
postional rms between the WIYN and MDM objects was 0.34 pixels which
at a scale of 0.1406189$^{\prime\prime}$/pix is roughly
0.05$^{\prime\prime}$.  Thirty six stars were used in the WIYN fit.
The error in the galaxy position was taken to be 0.05 pixel (as judged
from the scatter in the different position measurements) that
translates into a 0.007$^{\prime\prime}$ error in the lens position
(in each direction).   The Right Ascensions and Declinations of the
lensing galaxy and of the radio galaxy core and their relative
separation were measured using an average 
of three differences obtained from DAOFIND, IMEXAM and VISTA and 
are reproduced in Table\,\ref{tab:positions}.

\begin{table*}
\centering
\begin{tabular}{cllllr}
component & \mc{1}{c}{RA}             & \mc{1}{c}{Dec}            & \mc{1}{c}{east offset} & \mc{1}{c}{north offset} & \mc{1}{c}{flux density}\\
          & \mc{1}{c}{J2000}          & \mc{1}{c}{J2000}          & \mc{1}{c}{arcsec}      & \mc{1}{c}{arcsec} & \mc{1}{c}{mJy}\\
& & & & &\\
N2-E    &  08 16 38.82360 & +50 04 08.1275  & 13.5881  & 29.1493 &  1.12 $\pm$ 0.11\\
N2-W   &  08 16 38.81384 & +50 04 08.2098  & 13.4944  & 29.2316 & 0.93  $\pm$ 0.11\\
N1-E    &  08 16 38.86700 & +50 04 06.4285  & 14.0046  & 27.4503 &  4.01 $\pm$ 0.12\\
N1-W    & 08 16 38.82578 & +50 04 06.2481 & 13.6090 &  27.2699& 5.74 $\pm$ 0.12 \\
lens  &  08 16 38.82580 & +50 04 07.7950  & 13.6412  & 28.8530 & \\
\end{tabular}
\caption{\label{tab:positions} Fitted positions and flux densities of radio components.    }
\end{table*}

The positions of the pair (E and W) of doubled (N1 and N2) hotspot
components are listed in the first four rows of
Table\,\ref{tab:positions} and were determined using JMFIT within AIPS on
the image shown in Fig\,\ref{fig:squash}.  The positional uncertainty
along each axis is 4 milli-arcsec for the unresolved N2-E, N2-W and
N1-W components, and because of being slightly resolved that for the
N1-W component is 10 milli-arcsec.  The east and north offsets are
measured with respect to the position of the radio nucleus of the
radio galaxy (which is at 08 16 37.40774 +50 03 38.9782) which locks
the radio and optical frames together because of the coincidence of
the radio nucleus with the centre of the optical host galaxy.  The
positional uncertainty of the lensing galaxy is good to
$0.007^{\prime\prime}$ and dominated by the $0.5^{\prime}{\prime}$
seeing of the WIYN image.  The flux densities at 1.7\,GHz of the
compact lensed components are also measured using JMFIT and are for
the brightest compact features, excluding surrounding smooth extended
emission.

By applying a correction to the Right Ascension and Declination of the
radio galaxy to align it with the radio core, and applying this exact
same correction to the former RA and Dec exact relative offsets
between the lensed images and lens were then input to the model.

\section{Dimensionless model fitting}

The most northerly feature of the hotspot complex (N3) shown in
Figure~\ref{fig:vla_wht} is likely to be distorted (tangentially
sheared) by the presence of the lensing galaxy; it exhibits no
multiple compact components in the MERLIN data.  Thus, as is often the
case, the hotspot in this giant radio galaxy has multiple intrinsic
(unlensed) components \citep{Lai89}.  However, its impact parameter
from the galaxy means that it is not multiply imaged (see
Fig\,\ref{fig:caustic}).

\psfig{figure=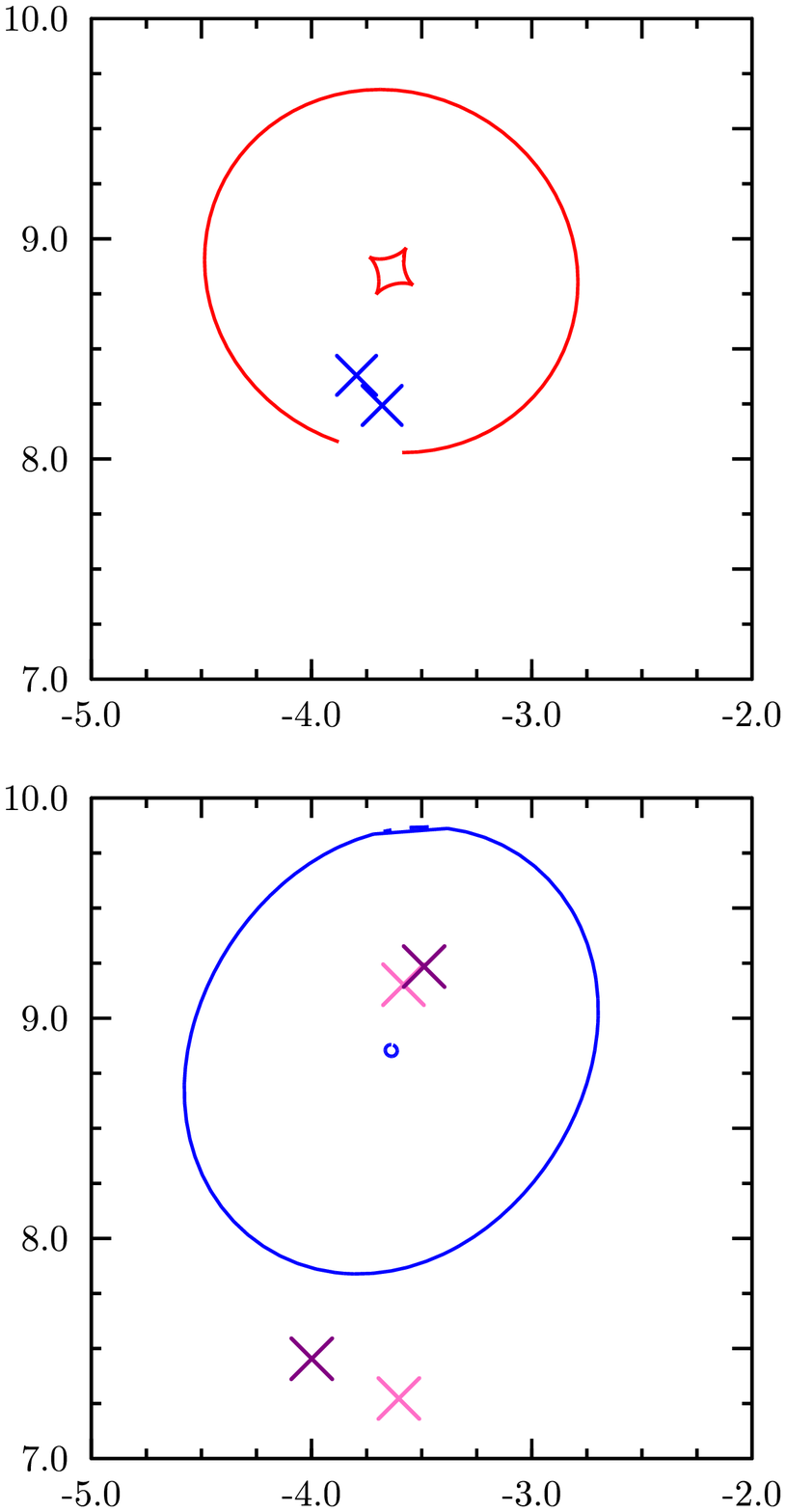,width=8cm} 
\figcaption{{\label{fig:caustic}
Upper panel: the source plane showing as blue crosses the positions of
the intrinsic double hotspot feature.   Lower panel:  image plane. }}

\subsection{General Approach}
\label{sec:generalapproach}

Gravitational lenses can be modelled in terms of a dimensionless
two-dimensional potential $\phi$ that is a function only of position
on the sky.  Dimensionful quantities -- distances and masses -- can
then be derived from redshift measurements and specification of the
Hubble constant.

While there is in principle a great deal of information to be
extracted from multiple images of extended sources, this quickly leads
into dividing the source into pixels, each of which is taken as a free
parameter \citep{Dye2005,Vegetti2009}.  Radio hotspots are
particular challenging because there is relatively little symmetry to
be exploited \citep{Lai89,Bla92,Har97,Lea97}.  
Moreover image infidelities may be present in the
details of the brightness structure.  For example, while the
resolution of the MERLIN image is approaching that ideally needed for
this system, the six antennas of the MERLIN array undersample much of
the extended structure seen in the VLA image, whose resolution is
insufficient to reveal the underlying structure.  Resolution of this
awaits the emerging generation of radio telescopes especially the
e-MERLIN and EVLA.  For the present we limit ourselves to modelling
the positions and fluxes of the compact components of the hotspots.

Reasonable and robust first-order results can be obtained using only
point sources.  We use the excellent public domain program {\em
  lensmodel} written and graciously made available by Chuck Keeton
\citep{Keeton2001}.\footnote{http://redfive.rutgers.edu/$\sim$keeton/gravlens/index.html}
Our sources are clearly extended, but we try to take out the
tangential stretching, seen clearly in the MERLIN image shown in
Figure\,\ref{fig:blobby}, by squashing by a factor of four in the
East-West direction.  The resulting image can be seen in
Figure\,\ref{fig:squash}.  Obtaining accurate radio positions of the
double images of each pair were made easier by the fact that one of
the knots shows no signal in Stokes $Q$ at the resolution of the
MERLIN data, so the position for the other pair was measured from a
Stokes $Q$ image.  To measure the other pair, the Stokes $Q$ image was
subtracted from a polarised intensity map $(\sqrt{Q^2 + U^2})$.  This
approach overcame the problem that the image shown in
Figure\,\ref{fig:blobby} does not completely resolve the two blobs in
N2.  [N.B., the rotation measure in the lensing galaxy implied by the
polarisation angle variation with frequency is only 20 rad/m$^2$,
modest by normal galaxy standards \citep{Lea86}.]

\subsection{Details of the model}
\label{sec:modeldetails}

We chose to model the two dimensional projected gravitational
potential with the function
\begin{equation}
\phi = \displaystyle b\left[\theta_x'^2 +
  \frac{\theta_y'^2}{(1-\epsilon)^2}\right]^{\alpha/2} 
\end{equation}
where $\theta_x'$ and $\theta_y'$ are angular position on the sky and
$\epsilon$ is the ellipticity of the equipotentials.  In the limit of
zero ellipticity, the radius of the Einstein ring, $r_{\rm E}$ can be
obtained from the parameter $b$ with $r_{\rm E} = b^{1/2-\alpha}$.
Potentials due to circularly symmetric isothermal galaxies (i.e.\ with
$\alpha = 1$) with external tides yield similar results but with
slightly inferior fits to the data.  For the small ellipticities
considered here there is no advantage to using elliptical isodensity
contours, which in the limit of small $\epsilon$ are a factor of 3
flatter than the equipotentials.

\psfig{figure=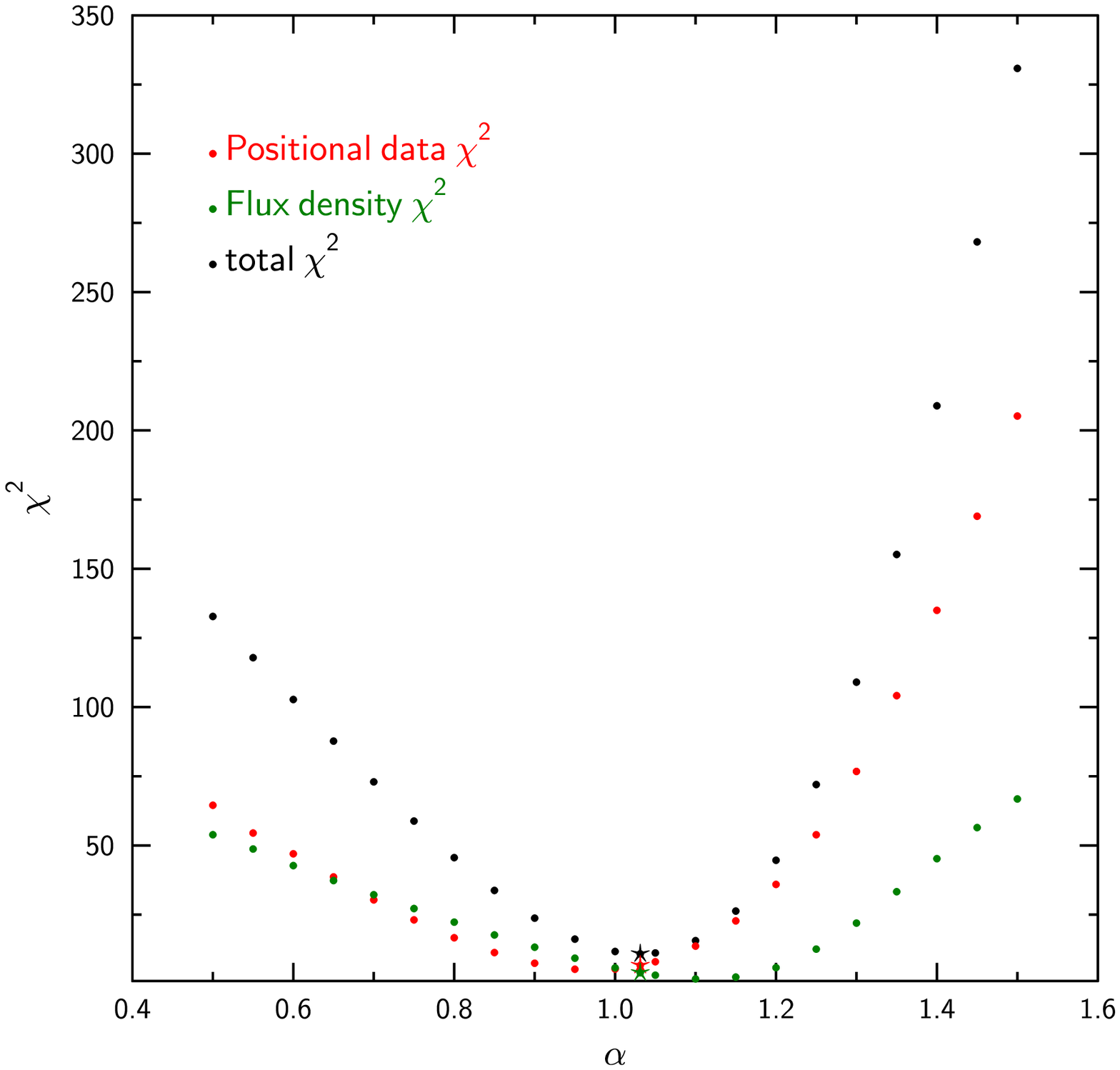,width=9cm,angle=0}
\figcaption{{\label{fig:alphachi} Plot of $\chi^2$ as a function of
    $\alpha$, the power-law slope of the potential.   }}

The results of stepping through $\alpha$ and allowing all the other 11 parameters vary are shown in
Figure\,\ref{fig:alphachi}.  Our best fit is for $\alpha = 1.031$, with an uncertainty, as judged from the change in $\chi^2$, of 0.002.  The best fitting model parameters are given in Table\,\ref{tab:paramtab}.  Note, in particular, that the best fitting position for the lens differs from our measured position by less than 0.003$^{\prime\prime}$.

\begin{table*}
\centering
\begin{tabular}{lc}
\mc{1}{c}{Parameter}              &                \\
lens strength $b$      &  0.906         \\
ellipticity $\epsilon$ &  0.061         \\
P.A.                   &  $57^\circ$     \\
lens offset (RA)       &  13.6389$^{\prime\prime}$     \\
lens offset (Dec)      &  28.8536$^{\prime\prime}$     \\
slope $\alpha$         &  1.031         \\
\end{tabular}
\caption{\label{tab:paramtab}  Lens parameters for the best fitting
model.  The position angle for the long axis of the equipotentials
is measured E from N.  In the limit of zero
ellipticity the radius of the Einstein ring, in arcseconds, is given 
by $r = b^{1/(2 - \alpha)}$.  }
\end{table*}

\section{Slope of lensing potential}
\label{sec:slope}

The use of a particular gravitational lens system in deducing $H_0$ is
only as good as our knowledge of the lensing potential.  The galaxy
responsible for lensing the hotspot in J\,0816+5003 has a power-law
which is remarkably close to isothermal.

In addition, it is unlikely that the lensing potential is much
shallower than isothermal because there is no hint of evidence for any
odd (e.g.\ `third') image \citep{Winn2003,Rus01}.  The flux density ratio of
the peak of the brighter northern (N2-E) blob on the MERLIN image to $3
\times$ the rms noise on this image exceeds 10.

\citet{Koc95}, similarly, finds that in MG\,1654+134 that $0.9 \leq
\alpha \leq 1.1$.  \citet{Rus02} find in B\,1152+199 that $0.95 <
\alpha < 1.21$ and \citet{Tre02} find in MG\,2016+112 that $\alpha =
1.0 \pm 0.1 \pm 0.1$.  However, seemingly inconsistently with this
emerging general picture, \citet{Che95} claim that the mass profile of
the lensing galaxy in MG\,1131+0456 to be incompatible with $\alpha =
1$.  In addition, for PG 1115+080 \citet{Rom99} report that no viable
lens model can yet be ruled out.

\section{Future work and conclusions}
\label{sec:conc}

Sensitive radio images at resolution better than a few 10s of
milli-arcsec would considerably broaden the scope of a lensing model
to be fitted, and this is within the capability of current
technology.  The key observational limitation however, remains the
registation of the optical frame with respect to the radio frame and
widespread efforts to improve this situation generally should remain a
priority.   

We conclude by remarking that our pure-lensing measurement of slope of
the potential of the lensing galaxy in this system as being very close
to isothermal is a measurement of the {\em total mass} on the scales
probed, including the dark matter.  However, it sheds no light on the
``bulge-halo conspiracy'' that separately the dark and light matter do
not appear to be isothermal --- the light matter being steeper than
isothermal, the dark matter being shallower
\citep{Treu2007,Bolton2007,Bolton2008,Gavazzi2007,Barnabe2009} --- but
collectively they appear isothermal over an increasing range in
redshift as the system we have studied in this paper illustrates.

\acknowledgments

PLS and KMB thank the Institute for Advanced Study at Princeton for
hospitality.  KMB thanks the Royal Society for a University Research
Fellowship.  PLS thanks the NSF for grants AST-0206010 and
AST-0607601.  We are very grateful to John Thorstensen for his
excellent astrometric software and graciously enabling us to get a
15-min image from the MDM-1.3\,m telescope and to Jules Halpern for
providing the image. We thank the staff of the MDM and WYIN
observatories.  MERLIN is a UK national facility operated by the
University of Manchester on behalf of STFC.  The UK Infrared Telescope
is operated by the Joint Astronomy Centre on behalf of the STFC. The
WHT is operated on the island of La Palma by the Isaac Newton Group in
the Spanish Observatorio del Roque de los Muchachos of the Instituto
de Astrofisica de Canarias.  The VLA is a facility of the NRAO
operated by Associated Universities, Inc., under co-operative
agreement with the National Science Foundation.  This research has
made use of the NASA/IPAC Extragalactic Database, which is operated by
the Jet Propulsion Laboratory, Caltech, under contract with the
National Aeronautics and Space Administration.   We thank the referee
for a careful reading of the manuscript.

\end{document}